\documentclass[]{aastex}
\usepackage{graphicx}
\usepackage{amssymb}
\usepackage{amsmath}
\usepackage{epstopdf}
\newcommand{\uo}{\mu_0}

\newcommand{\curl}{\nabla\times}

\newcommand{\diver}{\nabla\cdot}
\newcommand{\grad}{\nabla}

\newcommand{\db}{\mathbf{\delta B}}

\newcommand{\BB}{\mathbf{B}}

\newcommand{\VV}{\mathbf{V}}

\newcommand{\xxi}{\boldsymbol\xi}

\newcommand{\wa}{\omega_A}

\newcommand{\wm}{\omega_m}
\newcommand{\wt}{\omega^2}

\newcommand{\ddr}{\frac{d}{dr}}

\newcommand{\Af}{Alfv\'{e}n\ }
\newcommand{\beqn}[1]{\begin{equation} \label{#1}}
\newcommand{\eeqn}{\end{equation}}
\newcommand{\beqna}[1]{\begin{eqnarray} \label{#1}}
\newcommand{\eeqna}{\end{eqnarray}}

\title{Spectrum of Global Magnetorotational Instability in a Narrow Transition Layer}
\author{Jesse Pino \and S. M. Mahajan}
\affil{Institute for Fusion Studies, The University of Texas at Austin, Austin,Texas 78712}
\email{pino@mail.utexas.edu}
\begin{document}

\begin{abstract}
The Global Magnetorotational Instability (MRI) is investigated for a configuration in which the rotation frequency changes only in a narrow transition region. If the vertical wavelength of the unstable mode is of the same order or smaller than the width of this region, the growth rates can differ significantly from those given by a local analysis.  In addition, the non-axisymmetric spectrum admits overstable modes with a non-trivial dependence on azimuthal wavelength, a feature missed by the local theory. In the limit of vanishing transition region width, the Rayleigh-centrifugal instability is recovered in the axisymmetric case, and the Kelvin-Helmholtz instability in the non-axisymmetric case. 
 \end{abstract}
\maketitle

\section{\label{Intro}Introduction}

The Magnetorotational Instability (MRI), first discussed by \citet[]{Velikhov:1959} and  \citet[]{Chandrasekhar:1960}, became the subject of much study when it was proposed as an angular momentum transport mechanism in hydrodynamically stable astrophysical systems by Balbus and Hawley \citep[]{Balbus:1991,BH98,Hawley:1999}.  Much of the research on MRI was done using the local approximation, which uses the implicit assumption that the radial variation of the equilibrium flow is sufficiently small, allowing for modes to be sinusoidal in both the radial and axial directions.  However, this assumption can lead to misleading results about the existence \citep[]{2008ApJ...682..602M} and spectrum \citep[]{Pino:2008} of MRI modes. When the rotation profile changes significantly, globally unstable modes can be localized to the region of greatest shear. A recent study by \citet[]{mikhailovskii:052109} claimed to find `non-local' MRI modes in a configuration with a step-like velocity profile. However, this mode is actually a Rayleigh type surface mode, possessing the greatest growth rate in the hydrodynamic limit.  We show that a similar system with a narrow but finite region of velocity shear can support global MRI modes if the shear is stable to the Rayleigh centrifugal mode when the vertical magnetic field vanishes. These global modes are spatially confined to the transition region, and their properties differ from modes found using a local analysis.  In particular, there exists a discrete spectrum of global modes, with growth rates that differ significantly from the local prediction if the vertical wavelength is comparable to the transition region width. Overstability is possible for modes having an azimuthal wavelength component. Our investigations examine the full spectrum of such modes in both the axisymmetric and nonaxisymmetric case. 

\section{\label{BaseEquations} Base equations}
 We consider a generic rotating equilibrium in cylindrical geometry with flow $\VV = r \Omega(r) \hat{\theta}$ and mass density $\rho$, permeated by a uniform axial magnetic field $\BB = B_z \hat{z}$. We include gravitational and pressure terms in the ideal MHD equations, so that the system remains relevant to the global MRI in astrophysical settings such as accretion disks \citep[][]{FKR:2002,Hawley:2003} as well as to other rotating systems such as laboratory experiments \citep[e.g.][]{wang:102109,Goodman:2002} and stellar core collapse \citep[][]{2003ApJ...584..954A}. Since the equilibrium is uniform in the axial and azimuthal directions, we may take Fourier transforms in those directions (we neglect any vertical stratification or gravitational terms in what follows). The resulting equations for the normal modes of Lagrangian perturbations ($\xxi=\xxi(r)e^{ i(k_z z+m \theta -\omega t)} $) to this equilibrium are \citep[]{RevModPhys.32.898,1979MNRAS.187..769C,LBO67}:
\beqn{L-1}
-\omega^2 \rho \xxi - 2 i \rho\omega (\VV \cdot \grad)\xxi - \boldsymbol{\mathcal{F}}(\xxi) = 0 \ \ ,  
\eeqn
where
\beqna{L-2}
\boldsymbol{\cal{F}}(\xxi) = {}&   \grad(\gamma p \diver \xxi + (\xxi\cdot\grad)p) + \diver(\rho \xxi)\grad\Phi_g \nonumber \\  
 -{} & \ \grad(\BB\cdot \db)+(\BB \cdot \grad)\db +(\db \cdot \grad)\BB \nonumber  \\
 + {} &\ \diver(\rho \xxi (\VV \cdot \grad)\VV - \rho \VV  (\VV \cdot \grad)\xxi) \ \ .
\eeqna
 Here, $\gamma$ is the adiabatic index, and $p$ is the pressure, and $\Phi_g$ is the gravitational potential. We have ignored self-gravity effects by neglecting perturbations  $\delta \Phi_g$ to this potential. The magnetic field perturbation is $\db = \curl(\xxi \times \BB) $. The tensor divergence in the third line is taken with respect to the first coordinate, i.e. $\diver(\mathbf{AB}) =\nabla_i (A_i B_j)$.
 
 For incompressible perturbations  ($\diver\xxi = 0$), the radial component of Eq. (\ref{L-1}) and the divergence condition can be reduced to:
\beqna{mne0-1}
F_m \ddr \psi_T &=& \left[ F_m (F_m -\rho \frac{d  \Omega^2}{d \ln r} + \rho N^2)  - 4 \rho^2 \wm^2 \Omega^2 \right]\xi_r 
 + \frac{2 m \rho \wm \Omega}{r} \psi_T 
\eeqna
\beqn{mne0-2}
F_m\frac{1}{r} \ddr(r \xi_r) = -   \frac{2  m \rho \wm \Omega }{r} \xi_r + \frac{\zeta}{r^2} \psi_T
\eeqn
Where $\psi_T\ (=-(\xxi\cdot \nabla)p+\BB\cdot\db)$ is the total perturbed (gas + magnetic)  pressure, $\zeta = m^2+k_z^2 r^2$, $\wm =(\omega - m \Omega)$ is the Doppler-shifted frequency, $F_m=\rho(\wm^2 - k_z^2 v_A^2)$,  where $v_A= B_z/\sqrt{\rho\uo}$ is the \Af speed.  $N^2 = -(\rho'/\rho)(r \Omega^2- \partial_r \Phi_g)$ is the Brunt-V\"{a}is\"{a}l\"{a} frequency. Further reduction results in a single second order differential equation in $\xi_r$, the radial component of the Lagrangian perturbation.

\beqna{mne0eq1}
\left(\frac{r^3 F_m}{\zeta}\xi_r' \right)' &=
\biggl[& F_m \left(r  +\frac{\zeta}{r^2}-\frac{2m^2 r}{\zeta^2 }\right)   \nonumber\\   &-&  \frac{\rho k_z^2 r^3}{\zeta} \left(\frac{d  \Omega^2}{d \ln r} -  \frac{4 m  \wm \Omega}{\zeta}  
    + \frac{4 \rho \wm^2 \Omega^2}{ F_m}\right)  \nonumber \\
   &+& r \rho N^2-\frac{r^2 \rho'(\wt-m^2\Omega^2)}{\zeta} \biggr] \xi_r .
\eeqna
\subsection{Cartesian Limit}
This paper investigates global MRI when the rotation rate changes only over a small transition region of width $d$, centered at $r=r_0$.  When the region is narrow ($d \ll r_0$) and sufficiently far from the origin, it is expected that a Cartesian analysis should suffice to capture the essence of the mode.  Thus we drop any term which decays as $1/r$ or faster, except in the case of the  azimuthal mode number ($m/r$ is replaced by the continuous variable $k_y$). Choosing coordinates $x=r-r_0$, we take constant equilibrium rotation rates of $\Omega=\Omega_1$ for $x<-d$, and $\Omega=\Omega_2$ for $x>d$.  In what follows, we take a linear change in the rotation rate, i.e. 
\[\Omega(x) =\frac{\Omega_1+\Omega_2}{2} +  \frac{\Omega_2 - \Omega_1}{2}\frac{x}{d} = \bar{\Omega} + \frac{\Delta \Omega}{2}\frac{x}{d},\]
 although the analysis does not qualitatively change if different velocity profiles are considered. 
 In the frame rotating with the average angular velocity $\bar{\Omega}$, the Cartesian limit of Eq. (\ref{mne0eq1}) is:
\beqna{cartlimit}
\frac{d}{dx}\left( F_m \frac{d \xi_r}{dx} \right) =
\left[ F_m k^2+  k_z^2 \left(4 \bar{A}\bar{\Omega} \rho
     - \frac{4 \bar{\Omega}^2 \rho^2 \wm^2}{F_m}\right) - k_z^2\rho'(r \bar{\Omega}^2)+k^2 \rho' g \right] \xi_r.
\eeqna
Here, $k^2=k_y^2+k_z^2$, $\bar{A} = -1/2 \ d \Omega /d \ln r = -r_0 \Delta \Omega /(2 d) $ , and the Doppler shifted frequency in the transition region $|x|<d$ is $\wm=\omega+2 \bar{A} k_y x$. The total change in velocity across the transition region is $\Delta V=r_0\Delta {\Omega}= -4\bar{A}d$.  Exterior to the region, the rotation frequency and density are constant, and we have :
\beqn{exterior}
\xi_r'' = \left(k^2 - k_z^2\frac{4 \rho_{1,2}^2\bar{\Omega}^2 \omega_{1,2}^2}{F_{1,2}^2}\right)\xi_r,
\eeqn
with $\omega_{1,2}=\omega \pm 2\bar{A} k_y d$.  This admits decaying solutions of the form :
\beqn{kappa}
 \xi_{1,2} = \exp(-\kappa_{1,2} |x|) 
 \eeqn
\[ \kappa_{1,2} =  k \sqrt{1 -\frac{4k_z^2 \rho_{1,2}^2\bar{\Omega}^2 \omega_{1,2}^2}{k^2 F_{1,2}^2}}. \]
We must take the real part of the square root to be positive so as to ensure that the modes are bounded. The cartesian approximation taken here is similar to the shearing sheet model \citep[][]{1987MNRAS.228....1N,BH98}, however, our model has well defined boundaries which admit spatially decaying modes exterior to the transition region. 
\section{\label{Step}Step-like transition}
In this section, we show that the MRI is not present when Eq. (\ref{cartlimit}) is solved for a step-like transition, (i.e. the limit $d \to 0$).  \citet[][]{2004ApJ...601..414L} considered a similar situation in the context of QPO's generated by Kelvin-Helmholtz instability at an accretion disc-magnetosphere boundary layer. This limit was also taken by  \citet[][hereafter M08]{mikhailovskii:052109}, but resulted in the incorrect labeling of the instability as MRI. 
	Since the total pressure must be the same on both sides of the interface, the component of the Lagrangian perturbation normal to the interface, $\xi_r$, must be continuous \citep[]{LBO67}, but $\xi_r'$ can have a jump. Integrating across the narrow transition region, we arrive at: 
\beqn{fullstep}
F_2 \kappa_2 + F_1 \kappa_1 = \Delta \rho (k^2 g-k_z^2 r_0 \bar{\Omega}^2)-4 k_z^2 \bar{\rho} r_0 \bar{\Omega}(\Delta \Omega)
\eeqn
For axisymmetric modes,  $k=k_z$ and $\omega_{1,2}\to \omega$. Since the frequency then appears only through $\wt$, one can show that $\wt$ must be real \citep[see, e.g.][]{Chandrasekhar:1960}; therefore the frequency is purely real or purely imaginary. For an unstable root, we can take $\omega \to i \gamma, \ F_{1,2} \to -(\gamma^2 + k_z^2 v_A^2)$. For the remainder of the present paper, we will take the density constant across the boundary. Then Eq. (\ref{fullstep}) simplifies to: 
\beqn{axidisp2}
\sqrt{(\gamma^2+k_z^2v_A^2)^2 + 4 \bar{\Omega}^2 \gamma^2} = k_z^2 (r_0 (\Omega_1^2-\Omega_2^2)) 
 \eeqn
 This result is equivalent to the mode equation (26) of M08, in the limit of small rotation shear. We first note that instability is only possible for $\Omega_2 < \Omega_1$. Solving for $\gamma^2$, we find an unstable branch:
\beqn{g2}\gamma^2 = -(k_z^2 v_A^2+2 \bar{\Omega}^2) + 2\sqrt{\bar{\Omega}^2(\bar{\Omega}^2+k_z^2 v_A^2)+ k_z^2 r_0^2 (\Omega_1^2-\Omega_2^2)^2/4 }
\eeqn 
Taking the limit $v_A \to 0$, we arrive at the hydrodynamical result:
\[\gamma^2 = 2 \bar{\Omega}^2\left( \sqrt{1+ k_z^2 r_0^2 (\Omega_1-\Omega_2)^2/ \bar{\Omega}^4}-1\right)
\]
Near marginal stability, with $\gamma \ll \Omega$, we find $\gamma = k_z r_0 (\Omega_1-\Omega_2)/2= - k_z (\Delta V)/2$.  Furthermore, since $\gamma^2$ in Eq. (\ref{g2}) decreases with increasing $v_A$, the growth rate for a given wavelength is maximum when the field is absent. In the hydrodynamic limit, the instability exists at all wavelengths, the growth rate increases without bound. The addition of a vertical magnetic field acts as a surface tension, reducing the growth rate of all wavelengths, and stabilizing modes with wavenumbers above $k_{z,crit} = 2 \bar{\Omega} |\Delta V| /v_A^2$.  

In the fully azimuthal case $k_z=0$, $\kappa=k_y=k$, and Eq. (\ref{fullstep}) simplifies to: 
 \beqn{KH1}
 (\omega+k_y (\Delta V)/2)^2 + (\omega-k_y (\Delta V)/2))^2 = 0
 \eeqn
 This equation admits an imaginary solution with growth rate $\gamma =i \omega = k_y |\Delta V|/2$. This is the standard Kelvin-Helmholtz (KH) instability \citep[c.f.][\S 101]{Chandrasekhar:1961}.  We note that there is no centrifugal effect; the instability is due solely to the abrupt velocity change across the boundary. The growth rate of this instability is the same as for the axisymmetric Rayleigh instability if $k_z$ is replaced by $k_y$, however, the vertical magnetic field has no stabilizing effect in the pure azimuthal case.  When both $k_y$ and $k_z$ are nonzero, we again find that the magnetic field has only a stabilizing effect. If the density is not constant across the jump, additional destabilization can arise due to the Rayleigh-Taylor instability \citep[][]{2004ApJ...601..414L}.

In contrast to these results, the ideal magnetorotational instability occurs when a weak magnetic field destabilizes a plasma with a negative gradient in $\Omega$, because the frozen flux condition of MHD provides a coupling between adjacent fluid elements. For a given wavelength, there is a critical magnetic field strength which maximizes the growth rate of the MRI. This leads us to conclude that, contrary to the assertions in M08, treating a sharp velocity change in a rotating plasma as a step-like transition admits only magnetically stabilized hydrodynamical instabilities, but not the MRI. 

\section{\label{FiniteWidth} Finite-Width Transition}
 The step-like analysis that results in Eq. (\ref{fullstep}) fails to capture the MRI for two reasons. It makes the shear essentially infinite, and makes the implicit assumption that $\xi_r$ is constant throughout the transition region. A more robust treatment of the problem is to retain the finite transition region of width $2d$ and solve for $\xi_r$ on the interior. Normalizing lengths to $d$ and all frequencies to $\bar{\Omega}$, Eq. (\ref{cartlimit}) becomes: 
 \beqn{cartnorm}
\left( F_m \xi_r' \right)' =
\left[ F_m K^2+  K_z^2 \left(4 A
   - \frac{4 (\omega+2 K_y A x)^2}{F_m}\right) \right] \xi_r
\eeqn
Where $x$ runs from -1 to 1, $K = k d$ is the unitless wavenumber, and $A=\bar{A}/\bar{\Omega}$. In order to maintain consistency with the cartesian limit, we must maintain $d \ll r_0$. If $A= r_0(\Omega_2-\Omega_1)/(2 d \bar{\Omega})$ is not too large, we can still treat the whole system as rigidly rotating with a small additional shear flow. Corrections to this model are $O(A^2 d^2/r_0^2) \ll 1$.  
Global modes can now be found by matching $\xi$ and $\xi'$ at the left and right boundaries to the exterior solutions.       
\subsection{Axisymmetric modes}
If there is no azimuthal component to the mode wavelength ($K=K_z=k_z d$), the only position dependence is in the density. In the constant density case, the mode equation can be solved analytically,
\beqn{analyticmode}
 \xi_r'' = K_z^2 \left(1+\frac{4(1-A)(\gamma^2+\wa^2)-4\wa^2}{(\gamma^2+\wa^2)^2}
\right) \xi_r = -K_r^2 \xi_r.
\eeqn
Here $\wa=k_z v_A = K_z B_z/(\bar{\Omega}d\sqrt{\uo \rho})  $ is the normalized \Af frequency.  The eigenmode solutions are sinusoidal or exponential, depending on the sign of $K_r^2$. Both external boundary conditions can be simultaneously satisfied only when $K_r^2$ is positive. The local limit is achieved when $K_z \to \infty$, and gives an upper bound to the most unstable mode. Provided $A<1$, the maximum local growth rate is $\gamma = A$, occuring at an \Af frequency of $\wa = \sqrt{A(2-A)}$. These are characteristics of the local MRI \citep[]{BH98}. The instability is cut off for fields above $\wa = 2\sqrt{A}$ . We also see that for small magnetic field strength, the growth rate scales as $\gamma =\wa\sqrt{A/(1-A)} $. If $1<A<2$, the system is hydrodynamically unstable, but small magnetic field strengths will amplify the instability. Above $A=2$, the instability becomes the magnetically stabilized Rayleigh-centrifugal type, as in the previous section. 

For finite $K_z$, we can approximate solutions by taking rigid wall solutions ($\xi_r=0$) at the boundaries. Thus $K_r = n \pi/2$ for integer $n$. The eigenfunctions are real, and the mode number label denotes the number of times that solution crosses the origin. 
The system exhibits Sturmian behavior, that is, higher values of $n$ are more stable \citep[]{Goedbloed:2004}.  There is an unstable mode for every $n$ up to a critical $n_{max}$ which satisfies the marginal stability equation:
\beqn{nmax}
n_{max}^2  = \frac{4 K_z^2}{\pi^2}\left(4\frac{A}{\wa^2}-1\right) =\frac{4}{\pi^2}\left(4\frac{A d^2}{v_A^2}- K_z^2\right) .
\eeqn
When $K_z$ is large, there are many discrete modes which satisfy the global dispersion relation. The number of modes increases without bound as $K_z$ increases with fixed $\wa$. 
However, for fixed magnetic field strength $v_A$, the \Af frequency will also increase with $K_z$, reducing $n_{max}$, as well as the growth rate of the fastest growing mode. The critical parameter dictating the importance of these global effects is the ratio of the \Af transit time across the gap to the rotation rate, since for the fastest growing mode,  $\Omega d/v_A \sim K_z$. Even if the field is weak compared to the local rotation speed $v_A \ll r_0 \Omega$, global effects can be important in the $d \ll r_0$ limit. The number of unstable modes is reduced for small $K_z$.   Additionally, we have assumed infinite boundaries in the $z$ direction, but in real systems, the vertical system size $L_z$ imposes a minimum $K_z = 2 \pi d/L_z$.   
 
\begin{figure}
\centerline{
\includegraphics[width=\textwidth]{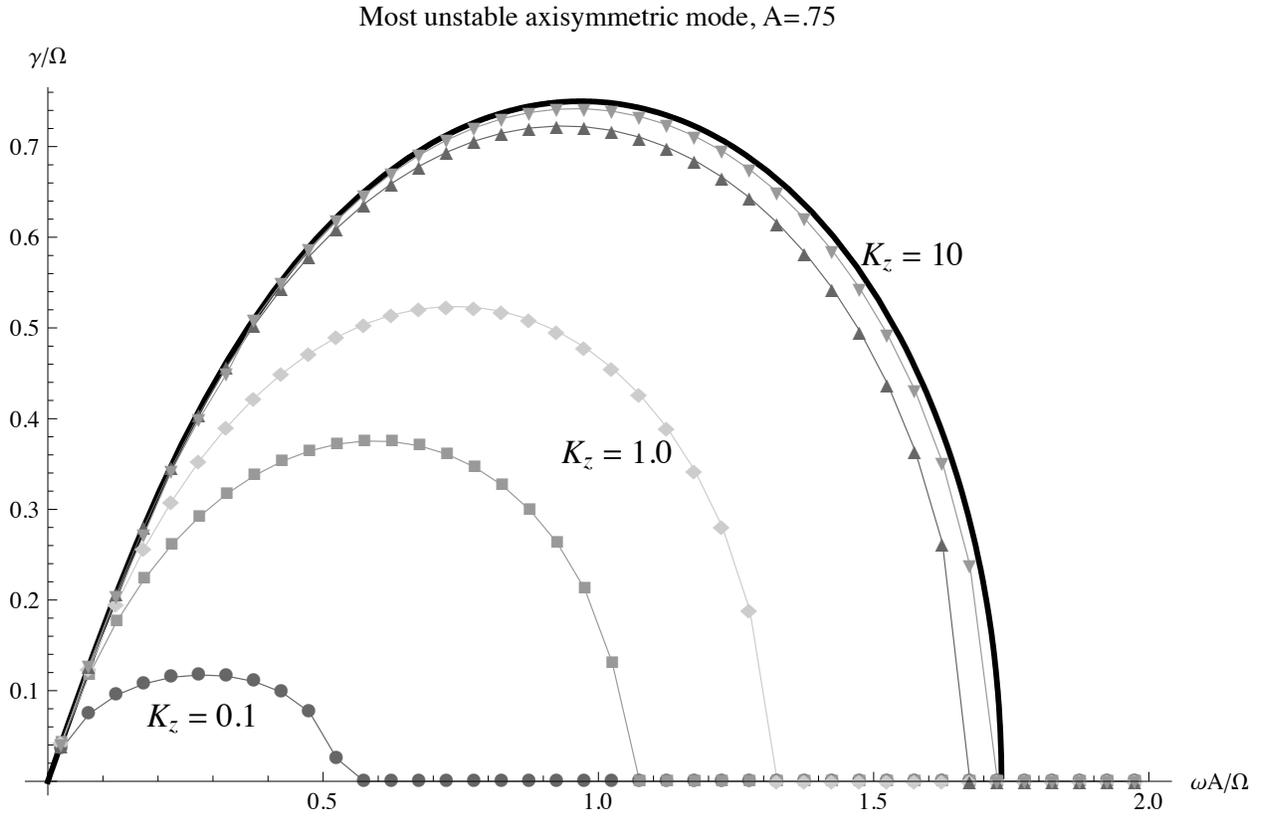}}
\caption{
\label{axilayer}
The most unstable mode (dotted lines) as a function of $\wa=k_z v_A$ for $A=0.75, K_z=k_z d = 0.1,0.5,1,5,10$. When $K_z$ is small, the maximum growth rate and the cutoff frequency are much lower than in the local limit (solid line).}
\end{figure}

\begin{figure}
\includegraphics[width=\textwidth]{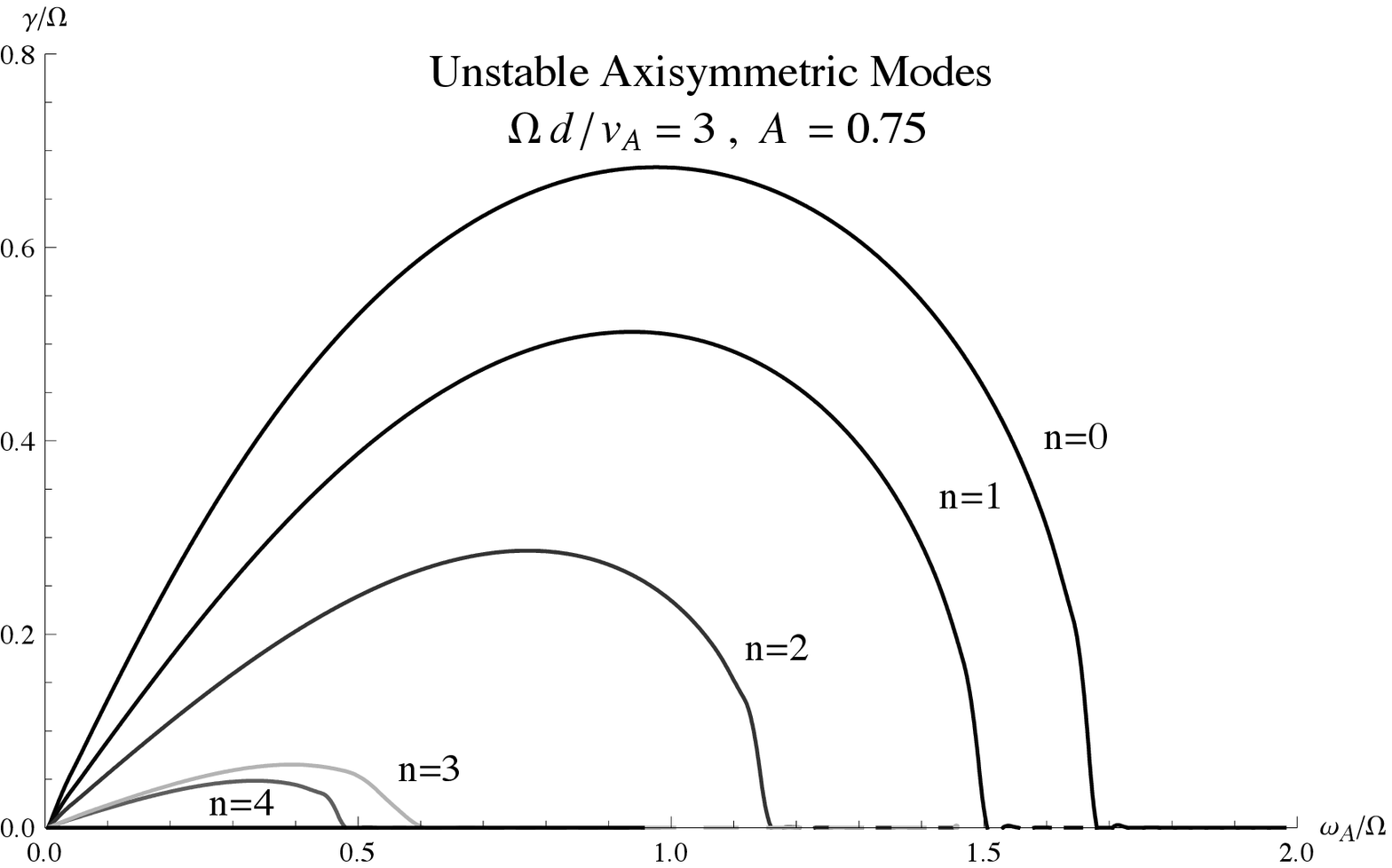}
\caption{
\label{aximodes}
Growth rates of all 5 global modes for $\Omega d/v_A = 3.0,\ A=0.75$. In this case $K_z = 3.0\  \wa/\Omega$. Global effects are more important for the smaller \Af frequencies.  }
\end{figure}

A shooting and matching code was used to find the discrete solutions that satisfy the boundary conditions that match Eq. (\ref{kappa}).  Figure \ref{axilayer} shows the most unstable global modes as a function of $\wa$ for fixed $K_z = 0.1,0.5,1,5,10$.  Note that the growth rate vanishes for vanishing magnetic field, one of the primary traits of the MRI. For smaller values of $k_z d$, the maximum growth rate and the cutoff frequency are much lower than in the local limit.  Figure \ref{aximodes} shows the growth rate of all unstable axisymmetric modes for when $\Omega d/v_A=3.0$. As $K_z$ (equivalently, $\wa$) is increased, the modes with smaller growth rate begin to be cut off, until there is only one mode remaining. 

\subsection{Nonaxisymmetric modes}
\subsubsection{Local theory}
The introduction of the azimuthal wavenumber removes the Hermiticity of the problem, and the frequency is no longer guaranteed to be purely real or imaginary, i.e., overstability is possible. However, if we take $K$ to be large with respect to the radial change in $\xi$, and restrict our analysis to small $x$, we can neglect this and obtain the local dispersion relation:  
\beqn{localky}
(\wt-\wa^2)^2+(4A(\wt-\wa^2)-4\wt) \cos^2 \theta
\
\eeqn
where $\cos \theta = K_z/K$. This yields a growth rate of
\[
\gamma^2 = -\wa^2 - 2(\Omega-A)\cos^2 \theta \pm 2 \Omega \cos \theta \sqrt{\wa^2+(\Omega-A)^2 \cos^2 \theta}, \] 
which has a maximum of $\gamma = A \cos\theta$, at $\wa =  \sqrt{A (2-A) \cos \theta}$. The instability is cut off above $\wa =  2 \sqrt{A} \cos \theta$. The only prediction of the local theory is that the addition of an azimuthal component of the wavelength is stabilizing compared to the axisymmetric case. 

\subsubsection{Global analysis}
The local approximation is a poor one in the non-axisymmetric case, as all dependence on position is neglected.  This approach fails to capture overstability, and sheds no light on the effect of finite $k_y$ on the spectrum of MRI. In particular, there is the introduction of a continuum of \Af singularities on the real frequency axis \citep[see, for example][for a discussion of the mathematical structure of this continuum]{HasegawaUberoi}. For every value of $-1<x<1$, there are two values of $\omega$ which satisfy $(\omega+2 A K_y x)^2 = \wa^2$. Thus there are two singularity regions, $\omega \in \{\pm \omega_A-2 A K_y, \pm \omega_A+2 A K_y\}$. 
When $\wa < 2 A K_y$, the two regions overlap. 

\begin{figure}
\includegraphics[width=\textwidth]{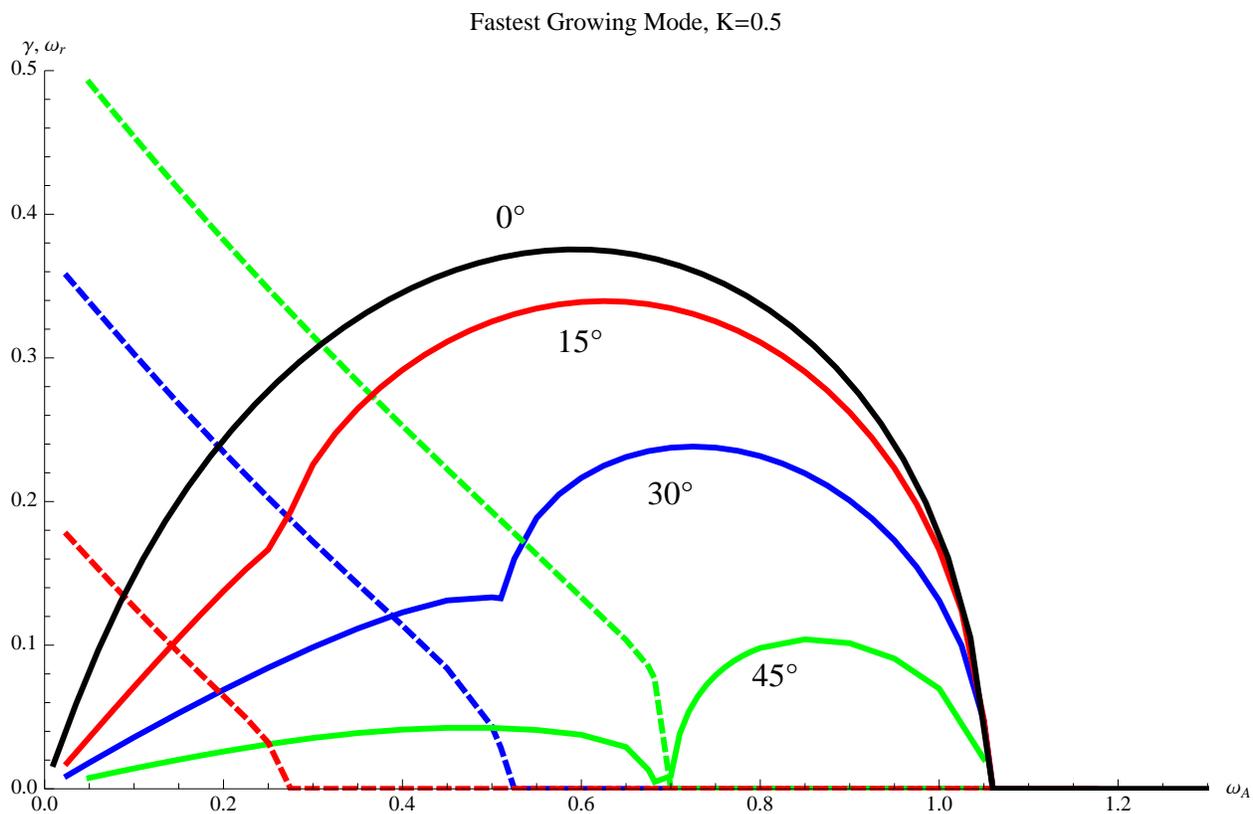}
\caption{\label{kp5} (Color online) Growth rates (solid) and real components (dashed) of the frequency for several angles ($\tan \theta = K_y/K_z$), with $K=0.5,\ A=0.75$. For smaller $\wa$, the real part approaches the \Af singularity value at the boundary, $\omega = 2 A K \sin \theta$, and the growth rate vanishes. For larger $\wa$, there are still pure imaginary modes which have the normal MRI dependence, but these decrease in growth rate as the angle is increased.}

\end{figure}

We have investigated the $K_y \ne 0$ mode equation (\ref{cartnorm})  with a shooting and matching code. All complex roots were found using a Lemur-Schur algorithm \citep[]{Acton:1997}.
Figure \ref{kp5} shows the dependence on the magnetic field strength of the most unstable axisymmetric modes for fixed $K_y$ and $K_z$, but at a variety of different pitch angles $\tan \theta = K_y/K_z$. We have taken a small value of $k d = 0.5$.  We first notice that the cutoff \Af frequency is the same for different angles, contrary to the predictions of the local analysis. When the magnetic field is low, the instability is suppressed compared to the axisymmetric case.  There is also a real part of the frequency, which indicates that the mode is co-rotating with the plasma at some point $x\ne 0$. As the magnetic field is decreased, this corotation point approaches the boundary, where $\omega_r =\pm 2 A K_y = \pm 2 A K \sin \theta$.  

\begin{figure}
\centerline{\includegraphics[width=\textwidth]{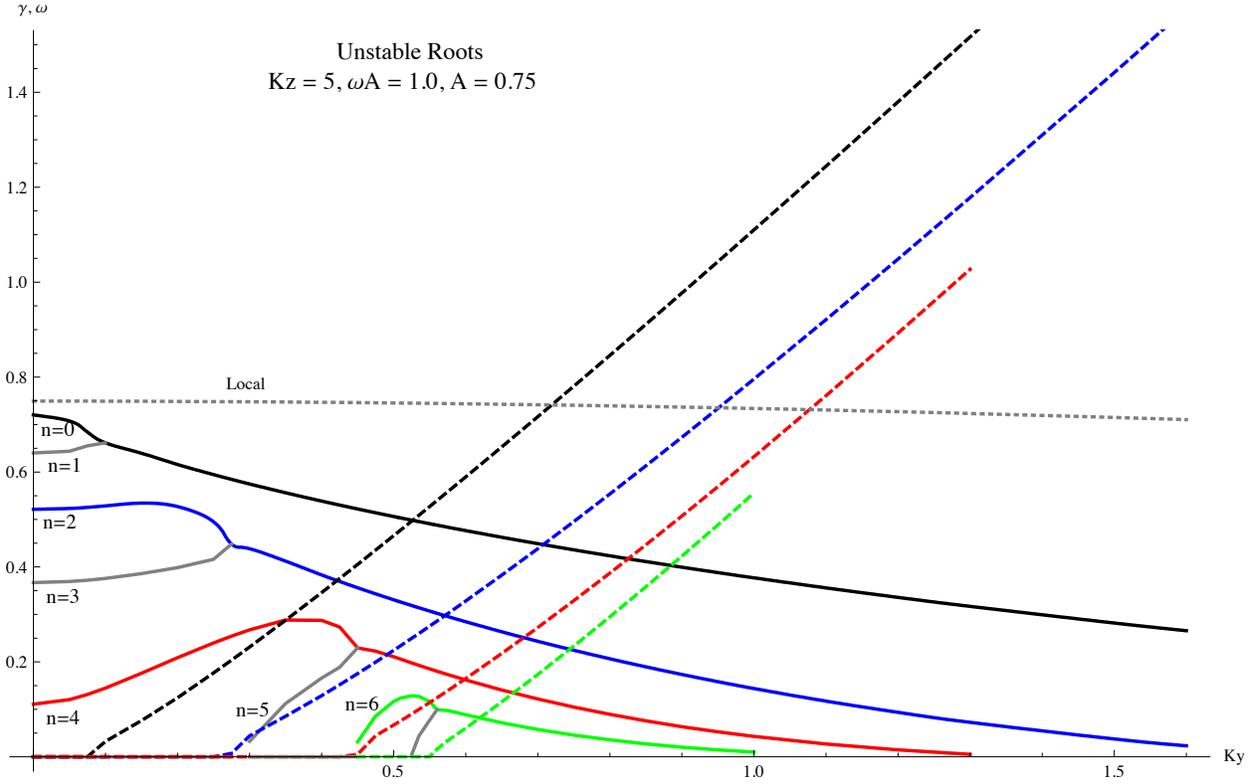}}
\caption{
\label{kz5wa1}
(Color online) All unstable global modes for $K_z = 5.0,\ A=0.75, \wa=1.0$. As $K_y$ is increased, pairs of purely growing modes become closer together in growth rate. At the point of merger, the frequencies have a real part (dashed lines) which starts at zero and increases linearly as $K_y$ increases. Near $K_y=0.28$, a new purely growing mode is created.  At a critical $K_y$ for each mode, the growth rate vanishes and the mode merges with the (stable) \Af continuum. The dotted line gives the maximum growth rate predicted by the local theory. Although the $n=0$ mode is close to this limit in the axisymmetric case, the global growth rates are significantly smaller for $K_y \ne 0$.}
\end{figure}

More can be learned by examining the structure of all unstable modes for increasing $K_y$. Figure \ref{kz5wa1} shows a representative example with fixed $K_z=5.0$ and $\wa=1.0$. In the axisymmetric limit, there are 5 purely growing modes for these parameters. As $K_y$ increases, the mode frequencies remain purely imaginary, although the eigenfunctions become complex (we retain the mode number ordering, although $n$ no longer refers to the number of nodes).  The growth rates of the some of the modes decrease, while those of others increase. At $K_y \sim 0.1$, the $n=0$ and $n=1$ modes have the same growth rate, and above that, both modes have the same growth rate, but they now have a real part. Due to the symmetry of the mode equation, if $\omega_r+i \omega_i$ are solutions, then so are  $\pm \omega_r\pm i \omega_i$. Here we only plot unstable roots with positive real part (dashed lines in Fig. \ref{kz5wa1}). As $K_y$ increases further, the real part increases linearly and the growth rate decreases. The growth rate of the most unstable global mode agrees fairly well with the local theory in the axisymmetric case, but the discrepancy grows rapidly as $K_y$ increases. 
 
At larger $K_y$, higher order mode pairs merge as well.  At $K_y \sim 0.28$, a new purely growing mode ($n=5$) becomes accessible, and eventually others are created as well. When $K_y\sim 1.0$,  the $n=6$ complex mode merges with the real \Af continuum. This trend continues until all instabilities are cut off above $K_y \sim 4.3$.

\subsection{Effective Potential Formulation}
\begin{figure}
\centerline{\includegraphics[width=\textwidth]{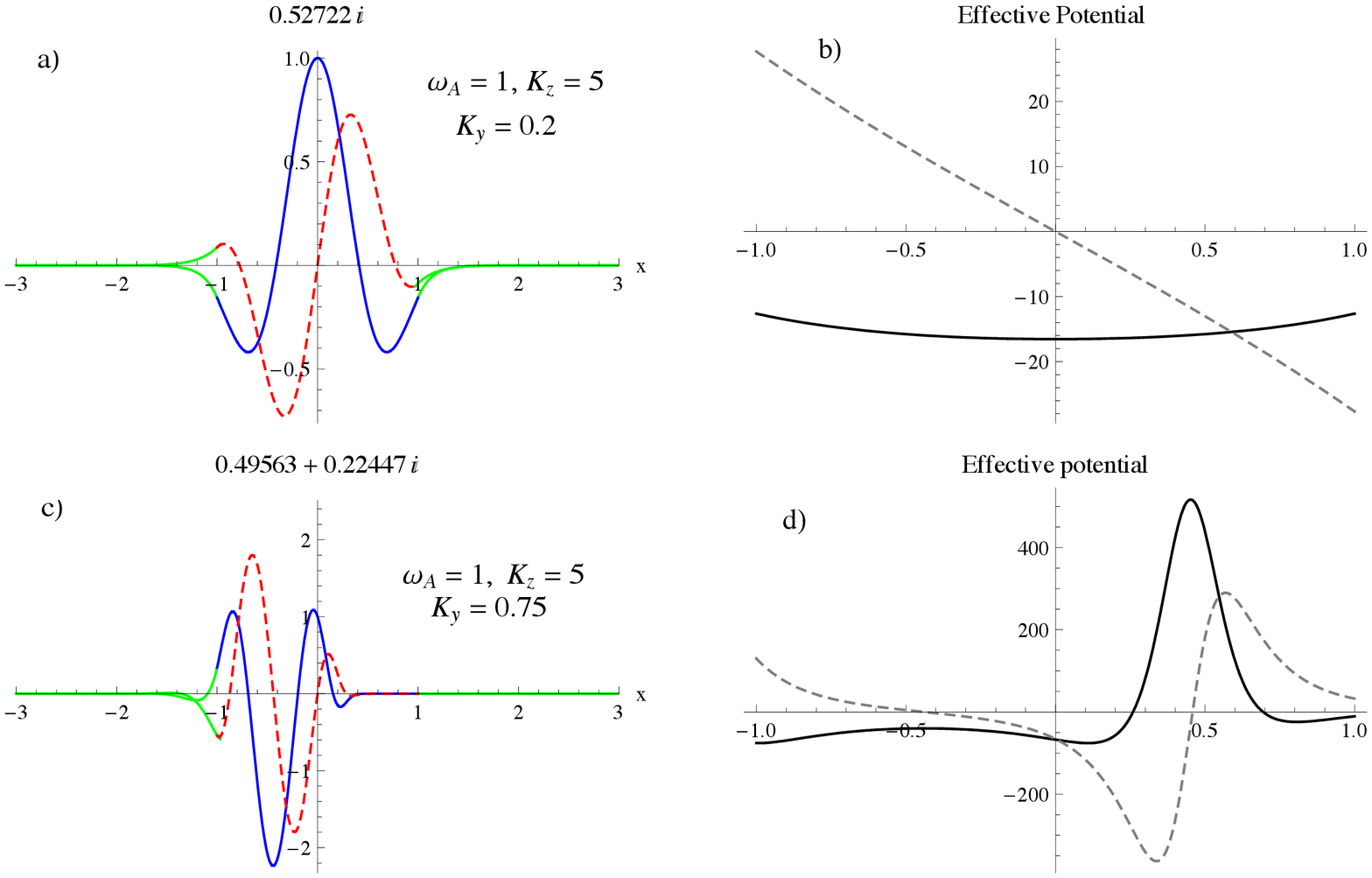}}
\caption{\label{potexplain}(Color online)  a) Eigenmode structure of $n=2$ mode ($\omega =0.5272 i$) for $K_z =5$, $K_y=0.2$, $\wa/\Omega=1.$, and $A=0.75$. The solid (dashed) line is the real (imaginary) part of $\xi_r$). b) Effective potential for the mode shown in (a). Since $K_y$ is small, $V$ is  changed little from the axisymmetric case of a negative real constant. c) For $K_y=0.75$, the same mode now has a complex frequency ($\omega = 0.4956+ 0.2245 i$), and has thus shifted its center to $x_0 = -\omega_r/(2 A ky)  = -0.472$. d) The effective potential of the mode shown in (c) displays the near-resonance barrier on the right. Since the mode has shifted to the left, the other peak is outside the transition region.}
\end{figure}

The behavior of the nonaxisymmetric modes for varying $K_y$ can be understood in the following manner. If we make the transformation $y=\sqrt{(\wm^2-\wa^2)}\xi_r$ then Eq. (\ref{cartnorm}) becomes:
\beqn{cartpot}
y''(x) = \left( K^2 - \frac{4 A^2 K_y^2 \wa^2 + 4 K_z^2 \wm^2 - 4 K_z^2 A (\wm^2-\wa^2)}{(\wm^2-\wa^2)^2}\right)y(x) = V(x;\omega) y(x),
\eeqn
recalling that $\wm = \omega + 2 A K_y x$. For a given set of parameters, there is an effective (complex) potential for the perturbation.  If the $y$ associated with a given potential can satisfy both boundary conditions, then that perturbation is an eigenmode of the system. The real part of the potential must therefore be sufficiently negative over enough of the transition region if the $|y|$ is to decay on both sides of the region (since $|y|'/|y|$ must change sign). Axisymmetric MRI eigenmodes have a $V$ which is a negative real constant over the region. Decreasing $\gamma = \Im(\omega)$ will make $V$ more negative, so smaller growth rates correspond to more spatial oscillations, in agreement with the previous section. In the nonaxisymmetric case, if $\omega$ is purely imaginary, $\Re(V)$ is symmetric about $x=0$ ($\Im(V)$ is antisymmetric). As long as $\gamma \ne 0$, there is no \Af singularity, but the potential changes the fastest near the points of closest approach, $x =(-\omega_r \pm \wa)/(2 A K_y)$. The real part of $V$ resembles a positive resonance curve near these points, with the growth rate $\gamma$ directly related to the width of the resonance. These `near-resonances' act as barriers, and the modes are evanescent inside them.  When $K_y$ is very small, these points lie far outside the transition region, and a given eigenmode remains similar to its axisymmetric counterpart, a seen in Figure \ref{potexplain}a. As $K_y$  increases, the resonances move closer to the boundary. Each mode at this point is related to either the left or right resonance, so it becomes necessary for the potential to shift, so as to keep only one resonance in the transition region (Figure \ref{potexplain}c). For an overstable mode, the $\Re(V)$ is symmetric around $x_0 = -\omega_r/(2 A K_y)$.  As the growth rate decreases and $\omega_r$ increases, the mode moves further over and becomes more oscillatory, approaching an \Af continuum mode at the boundary $x=\pm1$.      
 
\section{\label{Conclusions} Conclusions}

Global studies of the magnetorotational instability tend to depend strongly on specific boundary  conditions.  When the rotation rate changes only in a small area, however, velocity shear can give rise to an effective potential which spatially confines global eigenmodes and reduces dependence on the boundary conditions. We have examined an idealized case where $\Omega(r)$ changes sharply between two areas of rigid rotation; such a configuration has potential application to both astrophysical (e.g. \citet{2000ApJ...528..368H}, where sharp rotation gradients arise in models of stellar core collapse) as well laboratory settings (e.g. Ekman flow). When the velocity change is treated as a step-like transition as done in M08, the axisymmetric instability is most unstable in the limit of vanishing magnetic field-- we thus conclude that this is a Rayleigh type surface mode. The step-like transition is Kelvin-Helmholtz unstable as well in the non-axisymmetric regime. When the shear region has non-vanishing thickness, the unstable Rayleigh mode persists when $A= -1/2 d \ln\Omega d \ln r >1$. However, when the shear is moderate, ($A <1$) the flow is hydrodynamically stable, but is destabilized by the addition of a small vertical field.  This is the hallmark of the MRI. There is a spectrum of purely growing global axisymmetric instabilities.  If the vertical wavelength is comparable to the transition region width, the growth rates of the most unstable global modes differ significantly from the predictions of the local theory.

That the MRI results from the destabilization of the slow \Af wave by rotation was shown in \citet{BH98}. In the present analysis, we have shown that there is a spectrum of non-axisymmetric unstable and overstable global modes which connect smoothly to the \Af continuum. For a given axial wavenumber $k_z$, there is a complex dependence on the azimuthal wavenumber $k_y$. There exists a minimum azimuthal wavelength below which perturbations are stable. The interaction of these modes can have important consequences for turbulence in the nonlinear regime--this will be the subject of future work.   

The authors thank Dr. Richard Hazeltine and Dr. J. Craig Wheeler for useful 
discussions.

\end{document}